\documentclass[useAMS,usenatbib,letters]{mn2e}
\usepackage{aas_macros,graphicx,times,multirow}
\title[X-ray and optical observations of PSR J2144--3933]
{X-ray and optical observations of the closest isolated radio pulsar\thanks{Based on observations obtained with \emph{XMM-Newton}, an ESA science mission with instruments and contributions directly funded by ESA Member States and
  NASA, and with ESO/\vlt\ Antu (UT1).}}
\author[A.~Tiengo et al.]{A.~Tiengo,$^{1}$\thanks{E-mail: tiengo@iasf-milano.inaf.it}
R.P.~Mignani,$^{2}$ A.~De~Luca,$^{1,3}$ P.~Esposito,$^{4,5}$ A.~Pellizzoni$^{4}$ and S.~Mereghetti$^{1}$\smallskip\\
$^1$INAF - Istituto di Astrofisica Spaziale e Fisica Cosmica - Milano, via E.~Bassini 15, I-20133 Milano, Italy\\
$^2$University College London, Mullard Space Science Laboratory, Holmbury St. Mary, Dorking, Surrey RH5 6NT, UK\\
$^3$IUSS - Istituto Universitario di Studi Superiori, viale Lungo Ticino Sforza 56, I-27100 Pavia, Italy\\
$^4$INAF - Osservatorio Astronomico di Cagliari, localit{\`a} Poggio dei Pini, strada 54, I-09012 Capoterra, Italy\\
$^5$INFN - Istituto Nazionale di Fisica Nucleare, Sezione di Pavia, via A.~Bassi 6, I-27100 Pavia, Italy\\}
\date{Accepted 2010 December 26. Received 2010 December 20; in original form 2010 October 7}
\pagerange{\pageref{firstpage}--\pageref{lastpage}} \pubyear{2011}
\def\LaTeX{L\kern-.36em\raise.3ex\hbox{a}\kern-.15em
    T\kern-.1667em\lower.7ex\hbox{E}\kern-.125emX}

\def\xmm {{\em XMM-Newton}}
\def\vlt {{\em VLT}}

\def\psr {PSR\,J2144--3933}

\def\ltsima{$\; \buildrel < \over \sim \;$}
\def\lsim{\lower.5ex\hbox{\ltsima}}
\def\gtsima{$\; \buildrel > \over \sim \;$}
\def\gsim{\lower.5ex\hbox{\gtsima}}
\def\msole{~M_{\odot}}

\def\vlt{{\em VLT}}
\def\forsn{{\em FOcal Reducer/low dispersion Spectrograph}}
\def\fors{{\em FORS2}}

\begin{document}

\label{firstpage}

\maketitle

\begin{abstract}

With a parallactic distance of 170 pc, \psr\ is the closest isolated radio pulsar currently known. It is also the slowest ($P$ = 8.51 s) and least energetic ($\dot{E}_{\rm rot}=2.6\times10^{28}$ erg s$^{-1}$) radio pulsar; its radio emission is difficult to account for with standard pulsar models, since the position of \psr\ in the period--period derivative diagram is far beyond the typical radio `death lines'. Here we present the first deep X-ray and optical observations of \psr, performed in 2009 with \xmm\ and European Southern Observatory (ESO)/Very Large Telescope (VLT), from which we derive, assuming a blackbody emission spectrum, a surface temperature upper limit of $2.3\times10^5$ K for a 13 km radius neutron star, $4.4\times10^5$ K for a 500 m radius hot spot and $1.9\times10^6$ K for a 10 m radius polar cap. In addition, our non-detection of \psr\ constrains its non-thermal luminosity to be $<$30 per cent and $<$2 per cent of the pulsar rotational energy loss in the 0.5--2 keV X-ray band and in the $B$ optical band, respectively.


\end{abstract}

\section{Introduction}

The radio  pulsar \psr, discovered in the Parkes Southern Pulsar Survey \citep*{young99}, stands out among the nearly 2000 radio pulsars catalogued so far for several reasons. Its spin period ($P$) of 8.51 s is the longest of any known radio pulsars.
The rotation parameters of \psr\ are such that its
radio emission is surprising: its period and period derivative ($\dot{P}\simeq4\times10^{-16}$ s s$^{-1}$, corrected for the `Shklovskii effect'; \citealt{shklovskii70}) values locate \psr\ in the $P$--$\dot{P}$ diagram
below  most proposed `death lines' (e.g. \citealt{chen93}), where the accelerating potential is below the minimum value required to produce pair cascades and thus radio emission. Its radio emission has been explained either by proposing new models for pair production in the magnetosphere \citep{zhm00} or by suggesting that strong multipolar surface magnetic fields are present in all radio pulsars \citep{gil01}.

\psr\ has one of the largest characteristic ages ($\tau_c=P/(2\dot{P})\simeq3.4\times10^8$ years) among non-recycled radio pulsars and the lowest rotational energy loss ($\dot{E}_{\rm rot}=4\pi^2 I\dot{P}P^{-3}\simeq2.6\times10^{28}$ erg s$^{-1}$, where $I\approx10^{45}$ g cm$^2$ is the moment of inertia of the neutron star)
of any pulsar. The magnetic field, computed assuming
magnetodipolar spin-down, is
$B=\sqrt{\frac{3c^3I}{8\pi R^6}P\dot{P}}\simeq1.9\times10^{12}$
G, which instead is close to the average value for non-recycled radio pulsars.

A very faint object at radio wavelengths (apparent luminosity of $\sim$20 $\mu$Jy kpc$^2$ at 1400 MHz; \citealt{deller09}), \psr\ could be detected thanks to its small distance, only $\sim$180 pc as inferred from a dispersion measure of 3.35 cm$^{-3}$ pc assuming the Galactic free electron density model of \citet{taylor93}. This led \citet*{young99} to propose that \psr\ could be the tip of the iceberg of a huge population of Galactic long-period and low-luminosity pulsars, very difficult to detect but comparable in number to previous estimates of the total pulsar population of the Galaxy.
A much more robust distance of $165^{+17}_{-14}$ pc,\footnote{Correcting for the Lutz--Kelker bias \citep{lutz73}, the distance of \psr\ is only slightly increased to $172^{+20}_{-15}$ pc \citep*{verbiest10}.} based on the pulsar parallax, as well
as precise position ($<$1 mas) and proper motion measurement, $\mu_{\alpha}
= -57.89 \pm 0.88$  mas yr$^{-1}$ and $\mu_{\delta} = -155.9 \pm 0.54$ mas yr$^{-1}$, were
recently obtained
trough very long baseline interferometry (VLBI) observations at the Australian Long Baseline Array \citep{deller09}.
At the time of writing, according to the continuously updated on-line ATNF Pulsar Catalogue \citep{manchester05},\footnote{See http://www.atnf.csiro.au/research/pulsar/psrcat/} \psr\ is the closest isolated radio pulsar.


Despite its small distance, \psr\ has never been detected outside of the radio waveband.
Here we present the first deep optical and X-ray observations
of \psr, performed in 2009 with ESO/\vlt\ and \xmm.

\section{X-ray data analysis and results}



\xmm\ observed \psr\ for about 40 ks on 2009
October 24. Since the aim of the observation was the search for X-ray emission from the pulsar, only the data collected by the EPIC instrument, composed by three imaging detectors sensitive in the 0.2--15 keV energy range, were analysed.\footnote{\psr\ was also simultaneously observed with the Optical Monitor \citep{mason01short} for 36 ks with the UVW1 filter, but the observation was not sensitive enough to constrain significantly the pulsar emission.} The EPIC PN camera \citep{struder01short} was operated in full frame mode, while the two EPIC MOS units \citep{turner01short} were operated in large window mode. The thin optical blocking filter was used for the three EPIC cameras.
All the data were processed using the \xmm\ Science Analysis
Software (\textsc{sas} version 9.0.0). To improve the sensitivity at the lowest energy, where thermal emission from \psr\ is more likely to be detectable, the PN data were further processed using the \textsc{sas} tool \textsc{epreject},\footnote{See http://xmm.esac.esa.int/sas/current/doc/epreject/index.html.} which substantially reduces the low energy instrumental background due to electronic noise. The standard pattern selection criteria for
the EPIC X--ray events (patterns 0--4 for PN and 0--12 for MOS) were
adopted.  After filtering out the time intervals affected by a high level of particle background, the net exposure times were 22 ks for the PN and 26 ks for the MOS.

\begin{figure}
\centering
\resizebox{0.8\hsize}{!}{\includegraphics[angle=0]{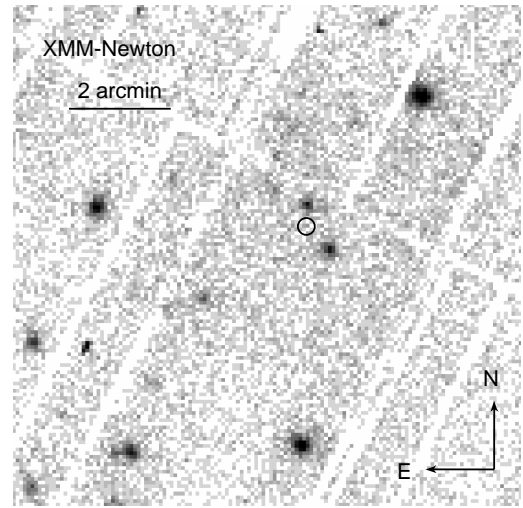}}
\caption{\label{ximage} \emph{XMM-Newton} EPIC image of the $\sim$10$\times$10 arcmin$^2$ field around \psr\ in the 0.2--1 keV energy range. The predicted position of the pulsar \citep{deller09}  is marked by a 10 arcsec radius circle. }
\end{figure}

A blind search for X-ray emission from \psr\ was performed using different source detection algorithms in many energy bands, but no source compatible with the pulsar position was detected. The closest X-ray source is
$25.6\pm0.5$ arcsec away from the proper-motion-corrected position of \psr\ (\citealt{deller09}, see Figure~\ref{ximage}), while the absolute astrometry of the EPIC images is $\sim$1 arcsec (root mean square, rms).\footnote{See http://xmm2.esac.esa.int/docs/documents/CAL-TN-0018.pdf.}

In order to set an upper limit on the pulsar X-ray emission, we selected the events from a 10 arcsec radius circular region centred at the position of \psr, obtaining, in the 0.2--10 keV energy range, 40 and 21 counts in the PN and in the two MOS cameras summed together, respectively  (see Table~\ref{xlimits} for other energy bands).
To better characterise the background intensity in different energy bands, we extracted for each camera the background events from a larger source-free region in the same chip, deriving the estimates reported in Table~\ref{xlimits} for
the number of background events expected in the source region.
Taking into account these background estimates and the Poissonian fluctuations \citep{gehrels86} of the counts extracted from the region of \psr, we obtain the 3$\sigma$ upper limits on the pulsar count rate in different energy bands reported in Table~\ref{xlimits}.
The conversion from count rates to fluxes for each detector is performed by assuming different spectral models (see Section 4) within the \textsc{xspec} (version 11.3.1) spectral fitting software and using
the response matrices and ancillary files produced by the \textsc{sas} software
for a 10 arcsec circular region centred at the pulsar position.



\begin{table}
\centering \caption{The 3$\sigma$ upper limits on the \psr\ count rates in different energy bands in the PN and the sum of the two MOS cameras (last column). These limits are derived from the number of events detected within 10 arcsec from \psr\ (third column) and the number of background counts expected in the same region (fourth column).
} \label{xlimits}
\begin{tabular}{@{}ccccc}
\hline
Energy band & Instrument & Observed & Background & 3 $\sigma$ upper limit\\
(keV) &  & counts & counts & (cts/s)\\
\hline
0.2--0.5 & PN & 5$\pm$2.2 & 6.8$\pm$0.5 & 4.9$\times$10$^{-4}$ \\
  & MOS & 5$\pm$2.2 & 2.4$\pm$0.1 & 5.4$\times$10$^{-4}$ \\
0.5--2  & PN & 15$\pm$3.9 & 10.1$\pm$0.6 & 1.0$\times$10$^{-3}$ \\
  & MOS & 8$\pm$2.8 & 7.4$\pm$0.2 & 5.4$\times$10$^{-4}$ \\
2--10  & PN & 20$\pm$4.5 & 16.7$\pm$0.8 & 1.1$\times$10$^{-3}$ \\
  & MOS & 8$\pm$2.8 & 8.4$\pm$0.2 & 5.0$\times$10$^{-4}$\\
0.2--10  & PN & 40$\pm$6.3 & 33.6$\pm$1.1 & 1.5$\times$10$^{-3}$ \\
  & MOS & 21$\pm$4.6 & 18.3$\pm$0.4 & 8.3$\times$10$^{-4}$ \\
\hline
\end{tabular}


\end{table}

\section{Optical data analysis and results}


We observed \psr\  with the \vlt\ at  the ESO  Paranal observatory on  2009 August 21
in visitor mode with  \forsn\ (\fors)
at the Cassegrain focus of the Antu UT1 telescope.
To enhance the sensitivity in  the blue part of the spectrum, we
used the \fors\  blue-sensitive CCD detector, a mosaic  of  two  $2k \times  4k$  E2V  CCDs  (with 15  $\mu$m  pixels) optimised  for wavelengths shorter  than 6000  \AA.  In  its standard resolution  mode,   the  detector  has  a  pixel   size  of  0.25 arcsec (2$\times$2 binning) which  corresponds to a projected field of view of 8$\farcm3  \times 8\farcm3$  over the CCD  mosaic.
To include a larger number of  reference stars for a precise  image astrometry and  photometry calibration,  as well  as to increase  the  signal-to-noise ratio  per  pixel,  we performed  the
observations in standard resolution mode.  We positioned the pulsar in the upper  CCD chip
to  exploit its larger  effective sky coverage  ($7 \times  4$ arcmin$^2$).   We chose  the standard  low
gain, fast read-out CCD mode.
We  observed   the  target  through   the  high-throughput  $U$ ($\lambda=3610$    \AA;    $\Delta   \lambda=505.1$ \AA),    $B$ ($\lambda=4400$  \AA;   $\Delta  \lambda=1035.1$ \AA),  and  $V$ ($\lambda=5570$ \AA; $\Delta  \lambda=1235$ \AA) filters.  A few bright stars
located relatively  close to the pulsar  position were  masked
to avoid ghost images and  saturation spikes.
We performed three  sequences of five  590 s  exposures each  for a  total integration  time of  8850 s through both  the $U$ and $B$ filters and a  sequence of  five 590 s  exposures through  the $V$ filter (2950 s integration time)
in  dark  time, photometric  sky conditions,
and
average air masses of $\sim$1.05,  $\sim$1.14, and $\sim$1.29 in the $U$,  $B$, and  $V$  filters, respectively.
The measured image quality (IQ) is $\sim 0.8$ arcsec in both the $U$ and $B$ filters
and
$\sim$$0.98$ arcsec in the $V$ one.

We  acquired bias  and twilight  flat--field frames  according  to the \fors\ science calibration plan
and images of
standard  star fields \citep{landolt92}
We reduced the science and standard star images
using the  last version of the           ESO            \fors\           data           reduction pipeline\footnote{See http://www.eso.org/observing/dfo/quality/FORS2/pipeline.} after producing master bias  and flat--field frames.  For each filter, reduced science images were stacked and averaged using available tools in {\em MIDAS}. We computed the photometric zero point using the \fors\ pipeline.
Unfortunately, since  the E2V detector is mounted at \fors\ only during visitor mode runs, no standard star observations are routinely taken to monitor the  stability of the zero point and to compute the  extinction coefficients in  different filters.
However,
we could use  as a reference the average  extinction coefficients computed for the E2V detector mounted  at {\em FORS1}.\footnote{See http://www.eso.org/observing/dfo/quality/FORS1/qc/qc1.html.}
We  estimate  that the  error  on  the zero  point  is $\sim$0.1 magnitude at most, likely  smaller than the expected photometric error expected  for a target  as faint  as the  pulsar.
%
%
%
%
As  a  reference for  the  \psr\ position  we  used  the proper-motion-corrected  VLBI coordinates  obtained  by  \citet{deller09}.
To accurately overlay  the pulsar coordinates, we have  then re-computed the astrometric solution of the \fors\ image,
using as a reference
objects selected  from the  Guide Star Catalogue  2 (GSC-2; \citealt{lasker08short}).
After accounting  for the  rms  of the astrometric  fit ($\sigma_{\rm  r} = 0.13$ arcsec),  the  uncertainty  of the registration of the \fors\ image on the  GSC-2  reference frame ($\sigma_{\rm tr}=0.13$ arcsec), and   the  0.15 arcsec uncertainty on the  link of the GSC-2  to the International  Celestial Reference Frame  (ICRF), we estimate that  the overall ($1 \sigma$) uncertainty  of our astrometry is $\delta_{r}=0.24$ arcsec. 

\begin{figure}
\centering
\resizebox{0.8\hsize}{!}{\includegraphics[angle=0]{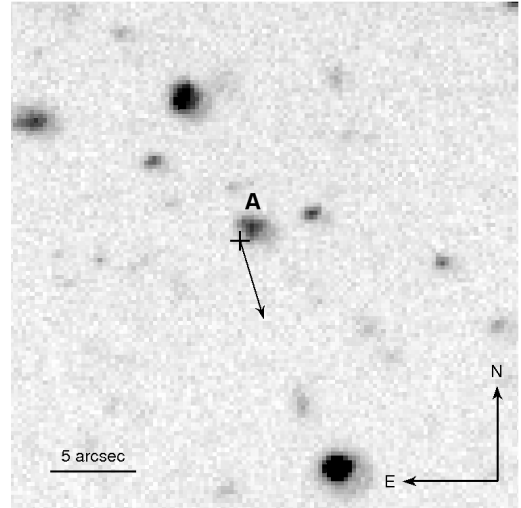}}
\caption{\label{oimage} ESO/\vlt\ B-filter image of the $\sim$30$\times$30 arcsec$^2$ field around \psr, whose predicted position is marked by a cross. The arrow indicates the proper motion direction and its length corresponds to the movement of \psr\ in 30 years \citep{deller09}. Object A, which is likely a background galaxy, is also indicated.}
\end{figure}


The position of \psr\ is  shown in Fig.~\ref{oimage}, overlaid on the co-added \fors\  $B$-band image.  As seen,  the pulsar position falls close to
a rather bright source which is  also detected in the $U$  and  $V$-band  images.   Its  coordinates,  $\alpha =21^{\rm  h}  44^{\rm  m}  11\fs97$,  $\delta  =  -39^\circ  33\arcmin 56\farcs42$,  imply an  offset of  $\sim 1.2$ arcsec  from  the computed pulsar position at the epoch  of our \vlt\ observations, i.e.  7 times larger  than  the overall  uncertainty  of  our astrometric  solution. Thus, we  can safely rule  out that this  object is associated  to the pulsar.    Its  magnitudes,   measured  through   customised  aperture photometry  with  \textsc{sextractor},  are  $U=23.39\pm  0.05$, $B=23.82 \pm 0.03$, $V=23.76 \pm 0.05$, after correcting for the air mass.   Its colours suggest that the  object (Object A) might be a background star, or, more likely,  a  galaxy, as suggested by its somewhat patchy morphology.
We could  not detect any other object  close  to the  pulsar  radio position  in  any  of the  \fors\ $U$, $B$, and $V$-band images.
The estimate  of the upper  limit is complicated by wings of Object A's  PSF which extend at the pulsar radio position. The possible extended nature of such object does not allow us to perform simple PSF subtraction. Thus, we used a different approach. We simulated a point source (modelled as a Gaussian, with a FWHM equal to the value measured for non-extended objects on our final coadded images), subtracted it at the expected pulsar position, and estimated the resulting residuals on the image in an aperture with a diameter equal to the FWHM. Such an exercise was repeated using increasing values for the magnitude of the simulated source, until residuals were consistent with a $3\sigma$ negative fluctuation. After correcting for atmospheric extinction, the resulting upper limits to the emission of PSR\, J2144$-$3933 are $U>25.3$, $B>26.6$, and $V>25.5$. 

\section{Discussion and conclusions}

\begin{figure}
\centering
\resizebox{0.9\hsize}{!}{\includegraphics[angle=0]{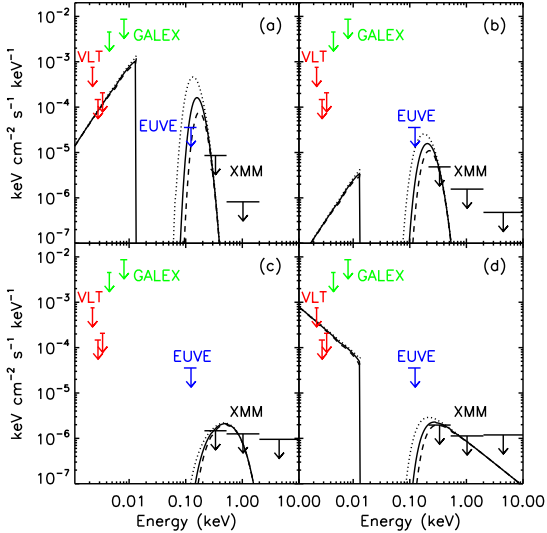}}
\caption{\label{sed} 3$\sigma$ upper limits
on the flux of \psr\ derived from our \vlt\ and \xmm\ observations and from archival {\it EUVE} and {\it GALEX} data.
The models used to derive the X-ray upper limits are shown for different values of interstellar extinction (\textsc{redden} model in \textsc{xspec}) and absorption (\textsc{phabs} in \textsc{xspec}) by the dotted [$E(B-V)=0.01$; $N_{\rm H}=5\times10^{19}$ cm$^{-2}$], solid [$E(B-V)=0.02$; $N_{\rm H}=10^{20}$ cm$^{-2}$] and dashed [$E(B-V)=0.03$; $N_{\rm H}=1.5\times10^{20}$ cm$^{-2}$] lines: blackbody with temperature $T=2.3\times10^5$ K and emission radius of 13 km (a), with temperature $T=4.4\times10^5$ K and 500 m radius (b),  with temperature $T=1.9\times10^6$ K and 10 m radius (c), and power-law with photon index $\Gamma=2$ and 0.5--2 keV luminosity of $7\times10^{27}$ erg s$^{-1}$ (d).}
\end{figure}

\psr\ being the closest isolated radio pulsar currently known, its non-detection in our deep \xmm\ and \vlt\ observations sets
robust upper limits on its X-ray and optical luminosity. Thanks to the high sensitivity of the \xmm\ PN instrument in the soft X-ray band, the best constraints on the thermal emission from \psr\ can be obtained from the PN data in the 0.2--0.5 keV band (see Table~\ref{xlimits} and Fig.~\ref{sed}).
Given the 170 pc distance, assuming a column density of $N_{\rm H}=10^{20}$ cm$^{-2}$ (as inferred from the radio dispersion measure for a 10 per cent ionization of the interstellar medium) and blackbody
emission from the whole surface of a 13 km radius neutron star,
the 3$\sigma$ upper limits on the temperature\footnote{The blackbody temperatures and radii reported here are the values measured at infinity. An observed radius of 13 km corresponds to an intrinsic radius of $\sim$10 km for a 1.4 $\msole$ neutron star.}
is $2.3\times10^5$ K ($kT=20$ eV).
However, the neutron star temperature is not expected to be uniform, with the magnetic poles hotter than the rest of the surface.
The radius of the polar cap can be estimated as $R_{\rm pc}=(2\pi R^3/cP)^{1/2}\sim 50$ m, where $R=10$ km is the neutron star radius and $P=8.51$ s is the pulse period. Significantly smaller emitting regions are expected, e.g., in the partially screened gap model \citep{gil08} and possibly observed in the thermal emission of old pulsars (e.g., \citealt{pavolv09}).
Assuming a 10 m radius polar cap, we obtain a temperature upper limit of $1.9\times10^6$ K ($kT=165$ eV) and a bolometric luminosity $L<$10$^{28}$ erg/s.
Our non-detection therefore implies a polar cap efficiency $L/\dot{E}_{\rm rot}<0.4$, not particularly constraining considering that efficiencies $<$1 per cent are typically observed in rotation-powered pulsars.
On the other hand, much larger hot spots are inferred from the X-ray spectra of isolated neutron stars possibly heated by magnetic field decay \citep{kaplan09}, which might be related to \psr\ (see below). Assuming a
500 m radius hot spot, we obtain a blackbody temperature upper limit of $4.4\times10^5$ K ($kT=38$ eV) and a bolometric luminosity $L<7\times10^{28}$ erg/s.

%

\psr\ was not detected in the extreme UV in a deep 20 ks observation with {\em EUVE} \citep{korpela98}. The flux upper limit of 0.023 $\mu$Jy
at 100 \AA\
is below the extrapolation to the extreme UV of the $T=2.3 \times 10^{5}$  K blackbody spectrum from the whole neutron star surface, if $N_{\rm H}=10^{20}$ cm$^{-2}$ is assumed. However, Fig.~\ref{sed} shows how the constraints derived from the {\em EUVE} data strongly depend on the interstellar absorption, while the presumably low absorption and extinction towards this nearby pulsar has no impact in the X-ray and optical band.
We also compared the extrapolation in the optical of the same blackbody spectrum with our {\em VLT}
flux upper  limits. Our deepest limit, obtained  through the  $B$  filter,
is a factor of $\approx  2$ above the Rayleigh-Jeans tail of the X-ray  blackbody spectrum, absorbed by an interstellar reddening $E(B-V)=0.02$.\footnote{This value for the reddening is computed using the  relation of \citet{predehl95} and assuming $N_{\rm H}=10^{20}$ cm$^{-2}$.}   Thus, it  does not
constrain the neutron   star  surface   temperature.
The \psr\ field was observed in direct image mode by {\em GALEX} \citep{martin05short} during the all-sky
survey for an exposure time of 208 s but no source was detected at the pulsar position down to a
$3 \sigma$ flux limit of $\sim 3$ $\mu$Jy and $\sim 6$ $\mu$Jy in the {\em GALEX} NUV ($\lambda=
 2771$ \AA; $\Delta \lambda=500$ \AA) and FUV band ($\lambda=
 1528$ \AA; $\Delta \lambda=200$ \AA), respectively. As can be seen in Fig.~\ref{sed}, these limits are well above the hottest blackbody compatible with the \xmm\ data.

Our limit of $2.3\times10^5$ K is among the lowest currently available
for the surface temperature of a neutron star.
Considering typical cooling models (see \citealt{yakovlev04} for a review) and the characteristic age of \psr\ (3.4$\times$10$^8$ years)
the neutron star surface temperature is expected to
be $<$10$^{5}$ K.
Possible evidence of surface temperature higher than expected from cooling models has been found for the 200 Myr
old PSR\, J0108$-$1431 \citep{mignani08,pavolv09,deller09}.
Moreover, the peculiar position of \psr\ in the pulsar period--period derivative diagram suggests a possible non-standard evolution and thus a more complex cooling history. For example,
this long period pulsar might be the descendant of a magnetar, i.e. a neutron star powered by the decay of its internal magnetic field of 10$^{14-15}$ G (see \citealt{mereghetti08} for a review). In this hypothesis,
the initially stronger magnetic field would have slowed down the neutron star cooling \citep{arras04} and, at the same time, the field decay would have heated up the neutron star surface either through Joule heating (e.g., \citealt{miralles98}) or possible magnetar-like bursting episodes. Moreover, in case of field decay, \psr\ would be much younger than its characteristic age, which largely overestimates the pulsar age when the magnetic field decreases with time.
%
This possibility is apparently precluded by the low dipolar magnetic field of \psr\ ($B=1.9\times10^{12}$ G), since numerical simulations of magnetic field decay in neutron stars have shown that the field decay should virtually stop at a field intensity of $\sim$2--3$\times$10$^{13}$ G \citep{pons09}. However, the recent discovery of a transient X-ray source with timing parameters similar to those of \psr\ ($P$=9.1 s, $\dot{P}$$<$$6\times10^{-15}$ s s$^{-1}$, and thus a dipolar magnetic field $B$$<$$7.5\times10^{12}$ G) showing magnetar-like activity indicates that a strong internal magnetic field might power the high energy emission of isolated neutron stars with dipolar magnetic fields well below 10$^{13}$ G \citep{rea2010}.
The X-ray dim isolated neutron stars (XDINSs, see \citealt{haberl07} for a review), which are nearby neutron stars with similarly long pulsation periods and surface temperatures of 5--12$\times$10$^5$ K, have also been suggested to be aged magnetars (see, e.g., \citealt{popov10}). When a period derivative has been measured, characteristic ages of 1--4$\times$10$^6$ years and dipolar magnetic fields of 1--3$\times$10$^{13}$ G have been derived. Although our non-detection of \psr\ cannot exclude the possibility that this radio pulsar was born as a magnetar, it indicates that, in this scenario, it would be older than the XDINSs, as already suggested by its lower dipolar magnetic field and longer characteristic age.

\psr\ being the radio pulsar with the lowest rotational energy loss, the search for non-thermal emission from this object is challenging as well. Assuming a power-law spectrum with photon index $\Gamma=2$ (see the lowermost panel of Figure~\ref{sed}), we can derive a 3$\sigma$ upper limit on the 0.5--2 keV observed flux of $1.9\times10^{-15}$ erg cm$^{-2}$ s$^{-1}$ from the PN data and $2.5\times10^{-15}$ erg cm$^{-2}$ s$^{-1}$ from the combined data of the two MOS units. The upper limit on the pulsar non-thermal luminosity in the 0.5--2 keV energy range is $7\times10^{27}$ erg s$^{-1}$, corresponding to $\sim$30 per cent of its  rotational energy loss. This limit is not particularly constraining, since the efficiency to convert pulsar spin-down power into X-ray luminosity is observed to be typically much lower \citep{liluli08}.
The  \vlt\  $B$-band
upper  limit  corresponds  to  a  luminosity $\la 5.6 \times 10^{26}$ erg s$^{-1}$, which is $\la$2 per cent of the spin-down  luminosity.  This is  about 4 orders of magnitude above the rotation-powered optical emission efficiency of pulsars with  comparable spin-down age.  For instance  the old pulsars PSR\,  B1929+10 ($\sim  3  \times 10^{6}$  years)  and PSR\,  B0950+08 ($\sim 17 \times 10^{6}$  years) have optical emission efficiencies of
$\approx 10^{-6}$ and $\approx  4 \times 10^{-6}$ \citep{zharikov06}, respectively.



%
%
%

To improve the  {\em XMM-Newton} temperature  upper limit  by  a  factor of  $\sim  2$, optical observations  deeper by  at  least $\sim  1.5$ magnitudes would  be required. Although  these observations are within reach of the \vlt, they are complicated by the presence of  Object A,  $\sim 1.2$ arcsec away  from the  pulsar position. This  would  require  an  image  quality much  better  than  0.8 arcsec, which is challenging  for ground-based optical  observations. A substantial improvement in sensitivity to surface thermal emission from \psr\ can instead be obtained with near-UV observations with the {\em HST}, which, apart from having a much better angular resolution, would take advantage from a more favorable waveband, around the peak of a $T$$<$10$^5$ K blackbody spectrum.


Considering that \psr\ is a nearby neutron star and that Object A is a relatively bright ($B$=23.8) background source (probably a galaxy), their small angular separation might cause a detectable gravitational lensing effect (see, e.g., \citealt{paczynski96hst}).
The displacement (or deformation, being the object slightly extended) of the background source due to the gravitational field of a point-like mass is expected to be \citep{paczynski96}

\begin{equation}
\delta \varphi =\frac{\varphi_E^2}{\Delta \varphi}, \label{lens1}
\end{equation}
where $\varphi_E$ is the Einstein ring radius and $\Delta \varphi$ is its angular distance from the lens. In the approximation of a very distant lensed object, the Einstein radius for a lens with mass $M$ at a distance $D$ is

\begin{equation}
\varphi_E =\bigg(\frac{4GM}{c^2 D}\bigg)^{1/2}. \label{lens2}
\end{equation}

Although the proximity of \psr\ ($D=170$ pc) and its relatively small angular separation from Object A ($\Delta \varphi=1.2$ arcsec) make it the most favorable case for any known neutron star,
for a mass of 1.4 $\msole$ a displacement of only $\delta \varphi=0.06$ mas is expected. Given the large and well constrained proper motion of \psr, two precise measurements of the galaxy apparent position in different epochs would be enough to measure its displacement and therefore to obtain a measure of the neutron star mass. Unfortunately, the astrometric precision of currently available optical instruments is not adequate for such a measurement. In the future, a sufficient angular resolution might be reachable, but this measure will not be possible anymore because the direction of the pulsar proper motion, which is opposite to the background object position and is not going to intercept any other bright source in the next decades (see Fig.~\ref{oimage}), will make the required precision rapidly increasing with time. %
Together with the other nearby neutron stars with a well determined proper motion, \psr\ is anyway a good candidate for the measure of a neutron star mass through microlensing (see, e.g., \citealt{dai10}). In fact, deeper observations, for example with the {\it HST} or the forthcoming {\it JWST}, might find dimmer sources along its sky trajectory, which, observed at their closest approach to the pulsar, might either show a
displacement or a flux increase. This measurements would be less challenging in the case of \psr\ than for the other nearby neutron stars, thanks to its vanishing flux outside the radio band.
%
%
%

\section*{acknowledgements}

We thank the referee, J. Gil, for his helpful comments and G. Mathys for his support in executing the \vlt\ observations.
We acknowledge the partial support from ASI (ASI/INAF contract I/088/06/0). PE acknowledges financial support from the Autonomous Region of Sardinia through a research grant under the program PO Sardegna FSE 2007--2013, L.R. 7/2007.

\bibliographystyle{mn2e}


\begin{thebibliography}{}

\bibitem[\protect\citeauthoryear{{Arras}, {Cumming} \& {Thompson}}{{Arras}
  et~al.}{2004}]{arras04}
{Arras} P.,  {Cumming} A.,    {Thompson} C.,  2004, \apjl, 608, L49

\bibitem[\protect\citeauthoryear{{Chen} \& {Ruderman}}{{Chen} \&
  {Ruderman}}{1993}]{chen93}
{Chen} K.,  {Ruderman} M.,  1993, \apj, 402, 264

\bibitem[\protect\citeauthoryear{{Dai}, {Xu} \& {Esamdin}}{{Dai}
  et~al.}{2010}]{dai10}
{Dai} S.,  {Xu} R.~X.,    {Esamdin} A.,  2010, \mnras, 405, 2754

\bibitem[\protect\citeauthoryear{{Deller}, {Tingay}, {Bailes} \&
  {Reynolds}}{{Deller} et~al.}{2009}]{deller09}
{Deller} A.~T.,  {Tingay} S.~J.,  {Bailes} M.,    {Reynolds} J.~E.,  2009,
  \apj, 701, 1243

\bibitem[\protect\citeauthoryear{{Gehrels}}{{Gehrels}}{1986}]{gehrels86}
{Gehrels} N.,  1986, \apj, 303, 336

\bibitem[\protect\citeauthoryear{{Gil}, {Haberl}, {Melikidze}, {Geppert},
  {Zhang} \& {Melikidze} Jr.}{{Gil} et~al.}{2008}]{gil08}
{Gil} J.,  {Haberl} F.,  {Melikidze} G.,  {Geppert} U.,  {Zhang} B.,
  {Melikidze} Jr. G.,  2008, \apj, 686, 497

\bibitem[\protect\citeauthoryear{{Gil} \& {Mitra}}{{Gil} \&
  {Mitra}}{2001}]{gil01}
{Gil} J.,  {Mitra} D.,  2001, \apj, 550, 38

\bibitem[\protect\citeauthoryear{{Haberl}}{{Haberl}}{2007}]{haberl07}
{Haberl} F.,  2007, \apss, 308, 181

\bibitem[\protect\citeauthoryear{{Kaplan} \& {van Kerkwijk}}{{Kaplan} \& {van
  Kerkwijk}}{2009}]{kaplan09}
{Kaplan} D.~L.,  {van Kerkwijk} M.~H.,  2009, \apj, 705, 798

\bibitem[\protect\citeauthoryear{{Korpela} \& {Bowyer}}{{Korpela} \&
  {Bowyer}}{1998}]{korpela98}
{Korpela} E.~J.,  {Bowyer} S.,  1998, \aj, 115, 2551

\bibitem[\protect\citeauthoryear{{Landolt}}{{Landolt}}{1992}]{landolt92}
{Landolt} A.~U.,  1992, \aj, 104, 340

\bibitem[\protect\citeauthoryear{{Lasker} et~al.,}{{Lasker}
  et~al.}{2008}]{lasker08short}
{Lasker} B.~M.,  et~al., 2008, \aj, 136, 735

\bibitem[\protect\citeauthoryear{{Li}, {Lu} \& {Li}}{{Li}
  et~al.}{2008}]{liluli08}
{Li} X.,  {Lu} F.,    {Li} Z.,  2008, \apj, 682, 1166

\bibitem[\protect\citeauthoryear{{Lutz} \& {Kelker}}{{Lutz} \&
  {Kelker}}{1973}]{lutz73}
{Lutz} T.~E.,  {Kelker} D.~H.,  1973, \pasp, 85, 573

\bibitem[\protect\citeauthoryear{{Manchester}, {Hobbs}, {Teoh} \&
  {Hobbs}}{{Manchester} et~al.}{2005}]{manchester05}
{Manchester} R.~N.,  {Hobbs} G.~B.,  {Teoh} A.,    {Hobbs} M.,  2005, \aj, 129,
  1993

\bibitem[\protect\citeauthoryear{{Martin} et~al.,}{{Martin}
  et~al.}{2005}]{martin05short}
{Martin} D.~C.,  et~al., 2005, \apjl, 619, L1

\bibitem[\protect\citeauthoryear{{Mason} et~al.,}{{Mason}
  et~al.}{2001}]{mason01short}
{Mason} K.~O.,  et~al., 2001, \aap, 365, L36

\bibitem[\protect\citeauthoryear{{Mereghetti}}{{Mereghetti}}{2008}]{mereghetti08}
{Mereghetti} S.,  2008, \aapr, 15, 225

\bibitem[\protect\citeauthoryear{{Mignani} et~al.,}{{Mignani}
  et~al.}{2008}]{mignani08}
{Mignani} R.~P., {Pavlov} G.~G., {Kargaltsev} O., 2008, \aap, 488, 1027

\bibitem[\protect\citeauthoryear{{Miralles}, {Urpin} \& {Konenkov}}{{Miralles}
  et~al.}{1998}]{miralles98}
{Miralles} J.~A.,  {Urpin} V.,    {Konenkov} D.,  1998, \apj, 503, 368

\bibitem[\protect\citeauthoryear{{Paczynski}}{{Paczynski}}{1996a}]{paczynski96}
{Paczynski} B.,  1996a, \araa, 34, 419

\bibitem[\protect\citeauthoryear{{Paczynski}}{{Paczynski}}{1996b}]{paczynski96%
hst}
{Paczynski} B.,  1996b, Acta Astronomica, 46, 291

\bibitem[\protect\citeauthoryear{{Pavlov}, {Kargaltsev}, {Wong} \&
  {Garmire}}{{Pavlov} et~al.}{2009}]{pavolv09}
{Pavlov} G.~G.,  {Kargaltsev} O.,  {Wong} J.~A.,    {Garmire} G.~P.,  2009,
  \apj, 691, 458

\bibitem[\protect\citeauthoryear{{Pons}, {Miralles} \& {Geppert}}{{Pons}
  et~al.}{2009}]{pons09}
{Pons} J.~A.,  {Miralles} J.~A.,    {Geppert} U.,  2009, \aap, 496, 207

\bibitem[\protect\citeauthoryear{{Popov}, {Pons}, {Miralles}, {Boldin} \&
  {Posselt}}{{Popov} et~al.}{2010}]{popov10}
{Popov} S.~B.,  {Pons} J.~A.,  {Miralles} J.~A.,  {Boldin} P.~A.,    {Posselt}
  B.,  2010, \mnras, 401, 2675

\bibitem[\protect\citeauthoryear{{Predehl} \& {Schmitt}}{{Predehl} \&
  {Schmitt}}{1995}]{predehl95}
{Predehl} P.,  {Schmitt} J.~H.~M.~M.,  1995, \aap, 293, 889

\bibitem[\protect\citeauthoryear{{Rea}, {Esposito}, {Turolla}, {Israel},
  {Zane}, {Stella}, {Mereghetti}, {Tiengo}, {G{\"o}tz}, {G{\"o}{\u g}{\"u}{\c
  s}} \& {Kouveliotou}}{{Rea} et~al.}{2010}]{rea2010}
{Rea} N.,  {Esposito} P.,  {Turolla} R.,  {Israel} G.~L.,  {Zane} S.,  {Stella}
  L.,  {Mereghetti} S.,  {Tiengo} A.,  {G{\"o}tz} D.,  {G{\"o}{\u g}{\"u}{\c
  s}} E.,    {Kouveliotou} C.,  2010, Science, 330, 944

\bibitem[\protect\citeauthoryear{{Shklovskii}}{{Shklovskii}}{1970}]{shklovskii%
70}
{Shklovskii} I.~S.,  1970, Soviet Astronomy, 13, 562

\bibitem[\protect\citeauthoryear{{Str{\"u}der} et~al.,}{{Str{\"u}der}
  et~al.}{2001}]{struder01short}
{Str{\"u}der} L.,  et~al., 2001, \aap, 365, L18

\bibitem[\protect\citeauthoryear{{Taylor} \& {Cordes}}{{Taylor} \&
  {Cordes}}{1993}]{taylor93}
{Taylor} J.~H.,  {Cordes} J.~M.,  1993, \apj, 411, 674

\bibitem[\protect\citeauthoryear{{Turner} et~al.,}{{Turner}
  et~al.}{2001}]{turner01short}
{Turner} M.~J.~L.,  et~al., 2001, \aap, 365, L27

\bibitem[\protect\citeauthoryear{{Verbiest}, {Lorimer} \&
  {McLaughlin}}{{Verbiest} et~al.}{2010}]{verbiest10}
{Verbiest} J.~P.~W.,  {Lorimer} D.~R.,    {McLaughlin} M.~A.,  2010, \mnras,
  405, 564

\bibitem[\protect\citeauthoryear{{Yakovlev} \& {Pethick}}{{Yakovlev} \&
  {Pethick}}{2004}]{yakovlev04}
{Yakovlev} D.~G.,  {Pethick} C.~J.,  2004, \araa, 42, 169

\bibitem[\protect\citeauthoryear{{Young}, {Manchester} \& {Johnston}}{{Young}
  et~al.}{1999}]{young99}
{Young} M.~D.,  {Manchester} R.~N.,    {Johnston} S.,  1999, \nat, 400, 848

\bibitem[\protect\citeauthoryear{{Zhang}, {Harding} \& {Muslimov}}{{Zhang}
  et~al.}{2000}]{zhm00}
{Zhang} B.,  {Harding} A.~K.,    {Muslimov} A.~G.,  2000, \apjl, 531, L135

\bibitem[\protect\citeauthoryear{{Zharikov}, {Shibanov} \&
  {Komarova}}{{Zharikov} et~al.}{2006}]{zharikov06}
{Zharikov} S.,  {Shibanov} Y.,    {Komarova} V.,  2006, Advances in Space
  Research, 37, 1979

\end{thebibliography}
\bsp

\label{lastpage}

\end{document}